# OPUS-Beta: A Statistical Potential for β-Sheet Contact Pattern in Proteins

Linglin Yu,[1] Mingyang Lu,[3] Tianwu Zang[1] & Jianpeng Ma[1, 2, 3, *]

[1]*Applied Physics Program and Department of Bioengineering,*
*Rice University*
*Houston, Texas, 77005*

[2]*Verna and Marrs McLean Department of Biochemistry and Molecular Biology,*
*Baylor College of Medicine,*
*Houston, Texas, 77030*

[3]*Center for Theoretical Biological Physics,*
*Rice University,*
*Houston, Texas, 77005*

[*]JM: To whom the correspondence should be addressed. Mailing address: One Baylor Plaza, BCM-125, Baylor College of Medicine, Houston, TX77030. Email: jpma@bcm.tmc.edu, Phone: 713-798-8187, Fax: 713-796-9438.

# Abstract

Developing an accurate scoring function is essential for successfully predicting protein structures. In this study, we developed a statistical potential function, called OPUS-Beta, for energetically evaluating β-sheet contact pattern (the entire residue-residue β-contacts of a protein) independent of the atomic coordinate information. The OPUS-Beta potential contains five terms, i.e., a self-packing term, a pairwise inter-strand packing term, a pairwise intra-strand packing term, a lattice term and a hydrogen-bonding term. The results show that, in recognizing the native β-contact pattern from decoys, OPUS-Beta potential outperforms the existing methods in literature, especially in combination with a method using 2D-recursive neural networks (about 5% and 23% improvements in top-1 and top-5 selections). We expect OPUS-Beta potential to be useful in β-sheet modeling for proteins.

# Introduction

Protein folding, whose goal is to determine three dimensional structures from one dimensional amino acid sequences, is a long-term challenging problem for both experimental and computational biophysics (Dobson and Karplus, 1999). Based on the assumption that the native structure has the lowest free energy among all the possible conformations of an amino acid sequence (Anfinsen, 1973), designing an effective scoring function is essential in accurately predicting the structures. Consequently, two types of potential functions are brought forward: physics-based potential functions (MacKerell *et al.*, 1998, Wang *et al.*, 2001, Ponder and Case, 2003, Wang *et al.*, 2004) and knowledge-based potential functions (Sippl, 1990, Sippl, 1995, Thomas and Dill, 1996, Lu and Skolnick, 2001, Lu *et al.*, 2003, Shen and Sali, 2006, Rykunov and Fiser, 2010). The physics-based potential functions are derived from the fundamental physics laws, while the knowledge-based potential functions are statistically calculated from the experimentally solved protein structures. Therefore, knowledge based potential function is also called statistical potential function. Limited as it is, statistical potential function practically surpasses physical potentials in many aspects (Bradley *et al.*, 2005, Skolnick, 2006), which mainly benefits from the explosively growth of experimental protein structures from the protein date bank (PDB) (Bernstein *et al.*, 1977, Koppensteiner and Sippl, 1998, Poole and Ranganathan, 2006, Mirzaie and Sadeghi, 2010, Deng *et al.*, 2012).

In protein structure prediction, one of the fundamental challenges is the prediction of β-sheet contacts, which requires accurate methods to evaluate and rank potential β-topologies or β-contact patterns for proteins. Although generic statistical potential functions (Zhou and Zhou, 2002, Shen and Sali, 2006, Lu *et al.*, 2008, Lu *et al.*, 2008, Zhang and Zhang, 2010, Zhou and Skolnick, 2011) have been developed, a specialized potential function for β-sheet modeling could be more efficient. In literature, potential functions based on the analysis of β-strand residue contact preferences have been reported. Hubbard (Hubbard, 1994) proposed a β-strand interaction pseudo-potential by scoring the interactions of inter-strand residue pairs. Hutchinson et al. (Hutchinson *et al.*, 1998) presented a detailed analysis of β-residue pair contact preference in antiparallel β-sheets of proteins according to hydrogen bonds in the antiparallel bridges. Zhu and Braun (Zhu and Braun, 1999) derived the pairwise contact map from the frequency of residue pairs in nearest, second and third contacts across neighboring β-strands. Steward and Thornton (Steward and Thornton, 2002) proposed an information theory approach based on the preference of a residue on a β-strand to be opposite a sequence of residues on an adjacent β-strand. Wu et al. (Wu *et al.*, 2007) developed the OPUS-Ca potential function containing a packing term in which there is an energy term specially designed for β-sheet structures. In addition, other computational methods such as support vector machines, recursive neural network, integer linear optimization, etc. (Cheng and Baldi, 2007, Tegge *et al.*, 2009, Rajgaria *et al.*, 2010, Savojardo *et al.*, 2013) were also brought forward for β-residue contact and β-sheet topology predictions. These methods provided valuable tools to get the residue-residue contact probabilities based on the sequence, from which pairwise pseudo energy function can be constructed for evaluating protein structures (Cheng and Baldi, 2005). However, a perfect energy function having the global minimum free energy in the native state of a protein is still waiting to be discovered.

In this paper, we present a new statistical potential, named OPUS-Beta, for energetically evaluating β-sheet contact pattern (the entire residue-residue β-contacts of a protein). To evaluate OPUS-Beta potential, no atomic coordinate information is required, and it only depends on the amino acid sequence and β-sheet contact pattern (the latter can be generated by other sampling methods or enumerations). OPUS-Beta potential contains five terms, including a self-packing term, a pairwise inter-strand packing term, a pairwise intra-strand packing term, a lattice term and a hydrogen-bonding term. We test the performance of OPUS-Beta in recognition of β-sheet topology and residue-residue contacts in β-sheets on several decoy-set collections specifically designed for this purpose. The results show that OPUS-Beta outperforms other methods in recognition of native β-contact pattern in both top-1 and top-5 selections. It is worth noting that the overall performance is dramatically improved by 5% and 23% for top-1 and top-5 selections respectively when OPUS-Beta is combined with the pairwise potential produced by 2D-recursive neural networks (2D-RNN) method (Cheng and Baldi, 2005, Tegge, Wang, Eickholt and Cheng, 2009). We expect this new potential to be helpful in improving the prediction of β-contacts in proteins.

## Materials and methods

**Structural Database**

First of all, a native structure library for non-redundant β proteins is constructed from the PISCES server (Wang and Dunbrack, 2003). We choose only X-ray determined structures with less than 25% identity, higher than 2.0 Å resolution, less than 0.3 R-factor, and more than 25% β-residues. To check the β-component of a structure, we use the Dictionary of Protein Secondary Structure (DSSP) (Kabsch and Sander, 1983) to assign secondary structures for the culled chains. To ensure the correctness of the assignment, we check the consistency between the PDB database and the DSSP database (Joosten *et al.*, 2011), and remove the assignment for the β-residues whose partners' paired residues are not themselves.

We further divide the structure library into three sets. The optimization set, which is used for optimization of the weights of the potential terms, contains 30 randomly chosen structures with more than 40% β-residues. The test set, which is used to generate a collection of decoy sets, contains another 100 randomly chosen structures with more than 40% β-residues and 8 to 14 β-strands. The rest 1205 native structures are grouped into the training set.

**Construction of OPUS-Beta**

The statistical potential has five terms: a self-packing term ($E_{self}$), a pairwise inter-strand packing term ($E_{pair\_inter}$), a pairwise intra-strand packing term ($E_{pair\_intra}$), a lattice term ($E_{lattice}$), and a hydrogen-bonding term ($E_{hydro}$):

$$E = w_{self} \sum_i E_{self}(A_i, L_i) + w_{pair\_inter} \sum_{ij} E_{pair\_inter}(A_i, A_j, L_{ij}) + \\ w_{pair\_intra} \sum_{ij} E_{pair\_intra}(A_i, A_j, L_{ij}) + w_{lattice} \sum_{ijkl} E_{lattice}(A_i, A_j, A_k, A_l) + w_{hydro} E_{hydro}.$$ (1)

Here, $w_{self}, w_{pair\_inter}, w_{pair\_intra}, w_{lattice}, w_{hydro}$ are the weights for the corresponding energy terms, $i, j, k, l$ are the residue indexes, $A$ represents the residue types, $L$ stands for the self-packing patterns in self-packing term or the pairwise interacting patterns in pairwise packing terms (see below for the definitions).

We also combine the potential function with the pseudoenergy for β-strand alignment in the 2D-RNN architecture (Cheng and Baldi, 2005, Tegge, Wang, Eickholt and Cheng, 2009, Fonseca *et al.*, 2010, Fonseca *et al.*, 2011, Subramani and Floudas, 2012). The combined statistical potential is a simple addition of the two potential functions, i.e.

$$E_{comb} = E + E_{2D-RNN}.$$ (2)

*Self-packing Term*

The self-packing term describes the preference of a residue type on a β-sheet. We specifically took into account different types of the β-strand on which the β-residues connect. The formula for obtaining the self-packing term is

$$E_{self}(A_i, L_i) = -k_B T \ln\left(\frac{N^{obs}(A_i, L_i)}{\langle N^{obs}(A_i, L_i)\rangle_{A_i}}\right), \tag{3}$$

where $N^{obs}(A_i, L_i)$ is the number of counts from the structural database for residue type $A_i$ at lattice state $L_i$, $\langle N^{obs}(A_i, L_i)\rangle_{A_i}$ is the reference state, which is the average value of observed counts over all possible residue types $A_i$. The state $L_i$ in sheet lattice is determined by the topologies of neighbor β-strands (Figure 1). Hence, there are a total of 5 states for $L_i$: one antiparallel partner, one parallel partner, two antiparallel partners, two parallel partners and two partners including one antiparallel partner and one parallel partner. Given that there are 20 different kinds of amino acids, we have $20 \times 5 = 100$ distinguished states for the self-packing energy term.

*Pairwise Packing Terms*

Presumably, the packing interactions between an inter-strand residue pair could be different from those between an intra-strand residue pair. Thus, we consider two different pairwise packing terms, each of which corresponds to the packing interactions of an inter-strand residue pair and an intra-strand residue pair respectively. These two terms have different weights in the whole OPUS-Beta potential.

The packing potential term for an inter-strand pair has the form of

$$E_{pair\_inter}(A_i, A_j, L_{ij}) = -k_B \ln\left(\frac{N^{obs}(A_i, A_j, L_{ij}) \sum_{A_i, A_j} N^{obs}(A_i, A_j, L_{ij})}{\sum_{A_i} N^{obs}(A_i, A_j, L_{ij}) \sum_{A_j} N^{obs}(A_i, A_j, L_{ij})}\right). \tag{4}$$

The packing potential term for an intra-strand pair has the form of

$$E_{pair\_intra}(A_i, A_j, L_{ij}) = -k_B T \ln\left(\frac{N^{obs}(A_i, A_j, L_{ij})}{\langle N^{obs}(A_i, A_j, L_{ij})\rangle_{A_i, A_j}}\right). \tag{5}$$

As the side chains of residues along a β-strand point upwards and downwards alternatively on a β-sheet, only the residues pairs whose side chains have the same direction are able to potentially contact with each other. Therefore, residue pairs $(A_i, A_j)$ are included in the statistical potential calculation only when their relative positions on β-strands and their side-chain orientation (upwards or downwards) allow side-chain packing interaction. Following the treatment from our previous study (Wu, Lu, Chen, Li and Ma, 2007), we consider all possible inter-strand contacting residue pairs. For the intra-strand pairs, we only consider the next nearest neighbor residues within the same strand. Consequently, the relative position $L_{ij}$ of the two residues in a β-sheet lattice is categorized into seven (four types for antiparallel strands and three types for parallel strands) types for inter-strand partners, and two types for intra-strand partners (Figure 2). The hydrogen bonds between the residues are determined by an improved hydrogen bond selection criteria (Fabiola *et al.*, 2002) during the analysis of the database.

*Lattice Term*

The lattice term describes the possible four-residue contact interactions within β-sheets. It involves the four residues in a square unit of the β-lattice, for example, $(i-1, i+1, j-1, j+1)$ in Figure 2. Once again, we mainly focus on the effects of the side chain interaction, so we count only the next nearest neighbors. The lattice term sets a global topology regulation of the β-sheets. Due to limit number of experimentally solved protein structures, we ignore the order of the different residues for obtaining a balance between the detailed description of β-sheet contact patterns and the number of possible energy states. The term has a form of

$$E_{lattice}(A_i, A_j, A_k, A_l) = -k_B T \ln\left(\frac{N^{obs}(A_i, A_j, A_k, A_l)}{\langle N^{obs}(A_i, A_j, A_k, A_l)\rangle_{i,j,k,l}}\right). \tag{6}$$

*Hydrogen-bonding Term*

The hydrogen-bonding term is set to be inversely proportional to the number of hydrogen bonds in the β-sheets. This term has no effects on the decoy recognition test described in the results section, but it could be effective in β-registrations modeling. The term has the form of

$$E_{hydro} = -N_{Hydro-Bond}. \tag{7}$$

**Construction of the Combined Potential**

The potential derived from the contact map produced by the 2D-RNN method is defined by

$$E_{2D-RNN}(P_{ij}) = -kT \ln(P_{ij}). \tag{8}$$

$P_{ij}$ is the contact possibility of residue $i,j$ in the protein sequence. Considering the size differences of the different protein structures, it is more reasonable to use the average value of this energy term in real calculations (Fonseca, Helles and Winter, 2010, Fonseca, Helles and Winter, 2011). Therefore, we make the energy term as below to have a fair combination with OPUS-Beta since the energy terms of OPUS-Beta are made from simple summations,

$$E_{2D-RNN} = \gamma \langle E_{2D-RNN}(P_{ij}) \rangle_{ij}. \tag{9}$$

Here $\gamma$ is a constant with an estimated value of 50, corresponding to the statistically estimated average number of β-residue pairs in β-proteins with 8 to 14 β-strands.

**Weights Optimization**

We optimize the weights of different energy terms via Monte Carlo sampling method against 30 protein structures in the optimization set. The PDB codes of these proteins are:

1k5n, 1o7i, 2ag4, 1lo7, 2zhp, 2wj5, 3bn0, 4gs3, 2gr8, 1nnx, 1itv, 1f00, 3cu9, 4h14, 1i4u, 1x8q, 4gai, 3pqh, 1pfb, 4h4n, 1d2s, 3zsj, 4hat, 1gwm, 3aeh, 3qs2, 1kqr, 2x4j, 1lyq, 2a15.

In each optimizing step, the weight for the self-packing term is fixed as 1.00, and the weight of each energy term is allowed to change by -0.25 to 0.25. After 800 cycles, the successful top-1 and top-5 selections of OPUS-Beta change from 0 and 1 to 0 and 6, and the optimized weights are

$$w_{self} = 1.0\ ;\ w_{pair-inter} = 2.1\ ;\ w_{pair-intra} = 0.7\ ;\ w_{lattice} = 0.5\ ;\ w_{hydro} = 0.7.$$

The decoy sets used in this optimization process are constructed by either shifting a specific β-strand forward or backward up to 3 residues or completely reversing the direction of that β-strand, or swapping a pair of β-strands in a β-sheet, which is elaborated further in the results section.

## Results

**Self-packing Energy**

In literature, tremendous attention has been paid to residue pair correlation in β-sheets (Hubbard, 1994, Hutchinson, Sessions, Thornton and Woolfson, 1998, Zhu and Braun, 1999, Steward and Thornton, 2002, Fooks *et al.*, 2006). It is also worthwhile to investigate the energetic preference of a particular amino-acid type involved in β-sheets (Caudron and Jestin, 2012). Figure 3 demonstrates the values of the self-packing energy of all 20 amino acids involved in five different packing patterns (see Methods section for details) in β-sheets. It appears that hydrophobic residues, VAL, LEU and ILE, are most favored by β-sheets in all packing patterns. Other residues, THR, ALA and PHE, are also favored by all packing patterns but with higher potential values comparing with the previous three. The last eight residues in Figure 3 (ASN, ASP, GLN, HIS, PRO, MET, CYS, TRP) display unfavorable self-packing energy in all packing patterns, and the rest of the residues (TRY, SER, GLY, LYS, ARG, GLU) have mixed energetic preference among the packing patterns. The statistical results are obtained from 1205 native protein structures, each of which has more than 25% β-component.

**Native Structure Recognition from Decoys**

To test the performance of the OPUS-Beta potential in recognizing native structures from structures with non-native β-sheet contacts, we specifically generated a collection of decoy sets, each of which contains a large sample of non-native β-contact configurations. The decoy set collection contains 100 different non-homologous β protein structures, each of which has 8 to 14 β-strands. Owing to the rich β contents of the test proteins, the decoy set collection covers a substantially large variation of β-contact patterns, allowing a thorough test on the effectiveness of a β-contact potential. We divided the decoy set collection into two groups according to the way we generated the decoys. In decoy set I, we generated decoys, for each protein, by shifting a specific β-strand (longer than two residues in length) along the direction of the strand forward or backward up to 3 residues or by completely reversing the direction of that β-strand. Therefore, for a protein with N eligible β-strands, there are $(3 \times 2 + 1)N$ decoys. Decoy set I is suitable for tests on the efficiency of a potential function in recognizing the right β registration. In decoy set II, we generated decoys, for each protein, by swapping a pair of β-strands in a β-sheet. If the two β-strands are of different lengths, we only swap the residues in the shorter one with part of the residues of the longer one near the amino terminus. As a result, we have $N(N-1)/2$ decoys for each protein. Decoy set II is suitable for tests on the efficiency of a potential in recognizing the right β-sheet topology.

In the decoy recognition tests, we compared the OPUS-Beta potential with a potential energy calculated by 2D-RNN method, and a combined potential with both the OPUS-Beta and the 2D-RNN methods. A random selection was also used as a control, by randomly selecting 1 or 5 β-contact patterns as native state in top-1 or top-5 test from the pool of native structure and the decoys of each protein.

The results for the tests on decoy set I are shown in Figure 4. In top-1 selection, for 7 out of 100 proteins, OPUS-Beta successfully distinguished the native β-contact pattern from the decoys. 2D-RNN achieved 6 out of 100, the random method achieved 2. More importantly, the combined potential by OPUS-Beta and 2D-RNN achieved 11. In top-5 selection, OPUS-Beta potential, 2D-NRR potential, the random method, and the combined potentials picked out 46, 28, 14, and 51, respectively. Dramatic improvement of decoy set recognition for the combined method suggests that the OPUS-Beta is orthogonal to the 2D-RNN potential, so that a combined usage of the two potentials has a superior performance over any one of them.

Tests were also performed on decoy set II. Figure 5 shows the results. In top-1 selection, OPUS-Beta has a similar performance with 2D-RNN; while in top-5 selection, OPUS-Beta is considerably better. The combined potential has the best performance (achieved 56 out of 100).

Figure 6 shows the results of the tests for each protein on a combined decoy set that contains decoys from both set I and II. Overall, the performance becomes worse for all methods since the size of each decoy set becomes larger. No obvious difference was observed for the performance between OPUS-Beta and 2D-RNN. However, the combined potential function has significantly better performance in both top-1 and top-5 selections.

**Discussions**

We developed a statistical potential (named OPUS-Beta) for energetically evaluating β-sheet contact pattern. In constructing the potential, we made use of the structural characteristics of β-sheet patterns in both β topology and registration level. These characteristics include self-packing pattern, pairwise packing interactions, hydrogen bonding arrangement, as well as amino acid group alignments, which correspond to the self-packing term, the pairwise packing term, the hydrogen-bonding term, and the lattice term, respectively.

A nontrivial problem in statistical potential construction is the assignment of weights to different energy terms (Feng *et al.*, 2007, Wu, Lu, Chen, Li and Ma, 2007). In our potential construction, weights optimization is processed by Mote Carlo sampling method to balance the effects of different energy terms. Among all the energy terms, the pairwise inter-strand packing term plays the most significant role followed by the self-packing term. In addition, the pairwise pseudoenegy produced by the 2D-RNN method, which can be derived from the one dimensional sequence, is combined with OPUS-Beta. Simple combination with 1:1 ratio of these two potential functions is used in our decoy tests.

Tests on recognition of native contact pattern from decoys show OPUS-Beta outperforms 2D-RNN potential in some specific cases, and is comparable with the later one in other aspects. A significantly enhanced performance (about 5% and 23% improvements) is observed for the combined potential function. Thus, the mutual enhancement between OPUS-Beta and 2D-RNN provides a basis for a better potential function in evaluating protein β-sheet contact patterns.

**Acknowledgements**

This work was supported by the National Institutes of Health Grant No. (GM067801) and Welch Grant No. (Q-1512). The authors acknowledge support from Blue BioU at Rice University. The computational software is available upon request.

# Figures

**Figure 1**

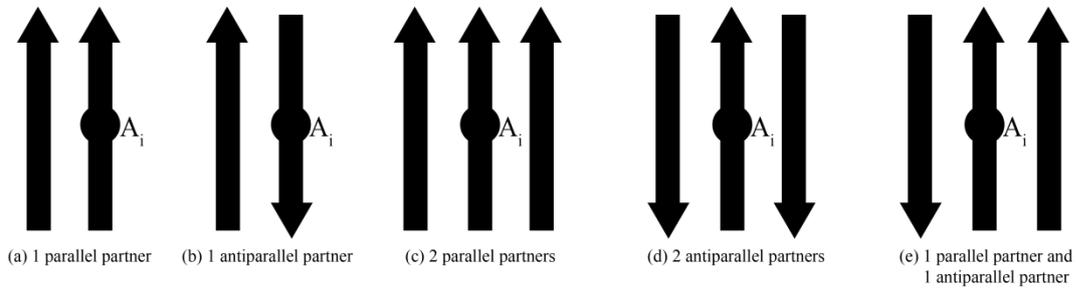

(a) 1 parallel partner  (b) 1 antiparallel partner  (c) 2 parallel partners  (d) 2 antiparallel partners  (e) 1 parallel partner and 1 antiparallel partner

**Figure 2**

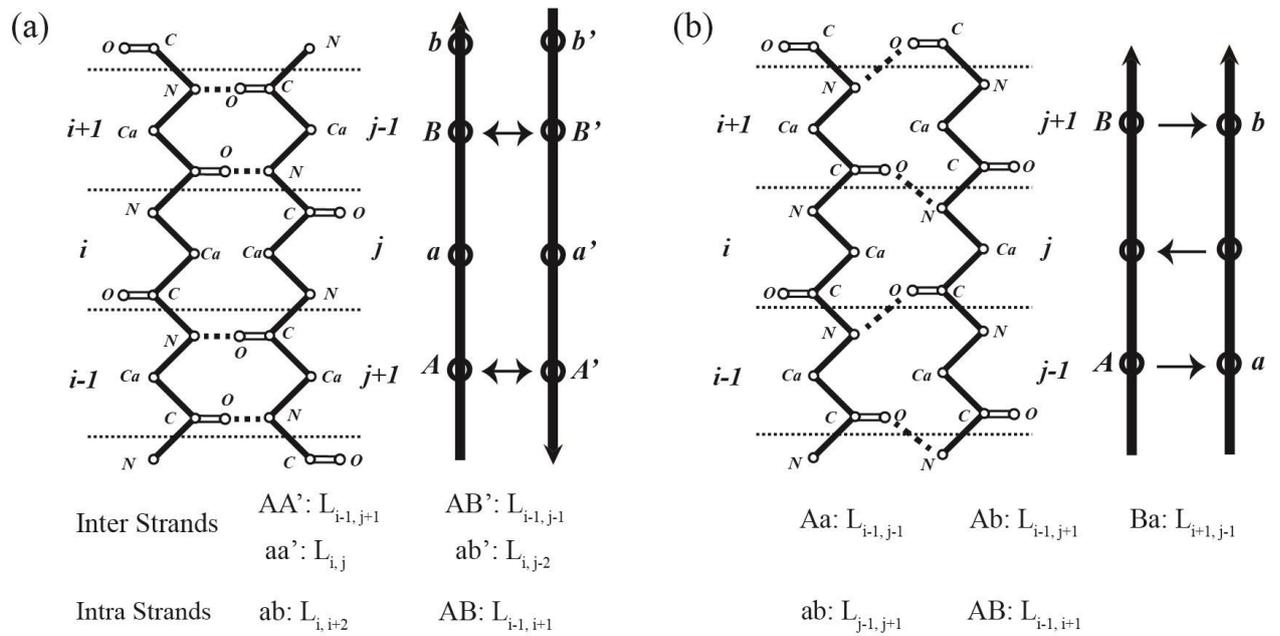

(a)

Inter Strands    AA': $L_{i-1,j+1}$    AB': $L_{i-1,j-1}$
              aa': $L_{i,j}$       ab': $L_{i,j-2}$

Intra Strands    ab: $L_{i,i+2}$      AB: $L_{i-1,i+1}$

(b)

Aa: $L_{i-1,j-1}$    Ab: $L_{i-1,j+1}$    Ba: $L_{i+1,j-1}$

ab: $L_{j-1,j+1}$    AB: $L_{i-1,i+1}$

**Figure 3**

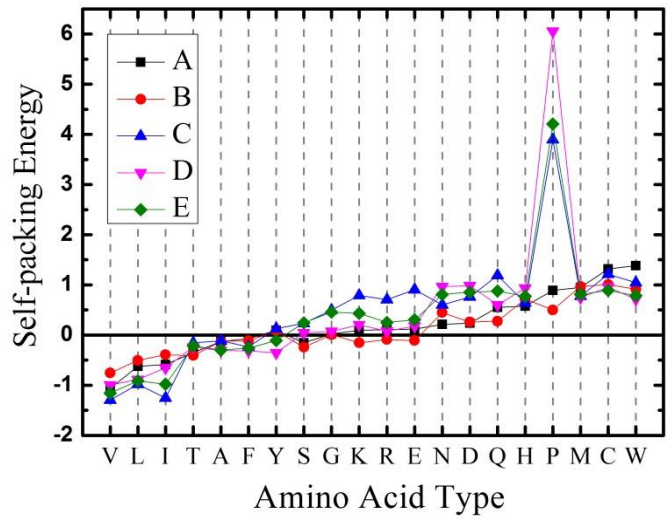

**Figure 4**

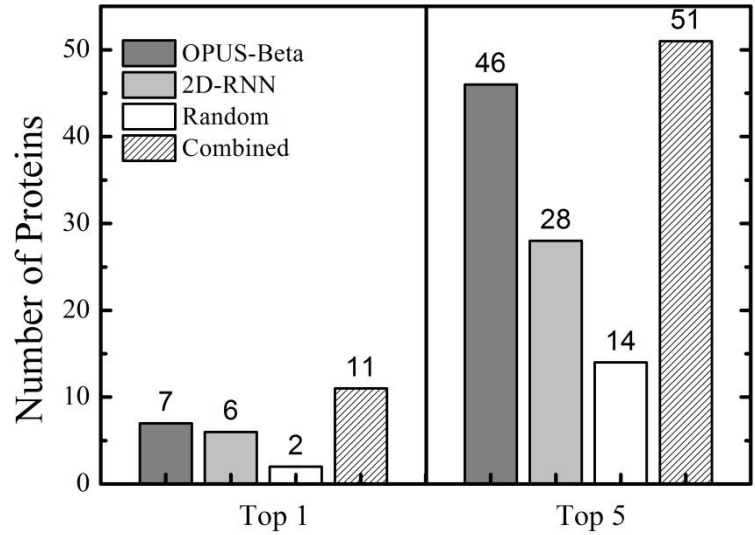

**Figure 5**

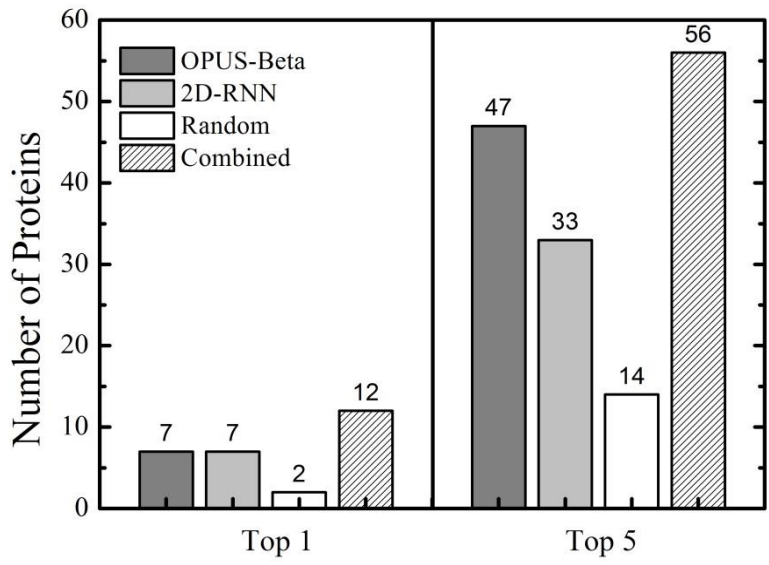

**Figure 6**

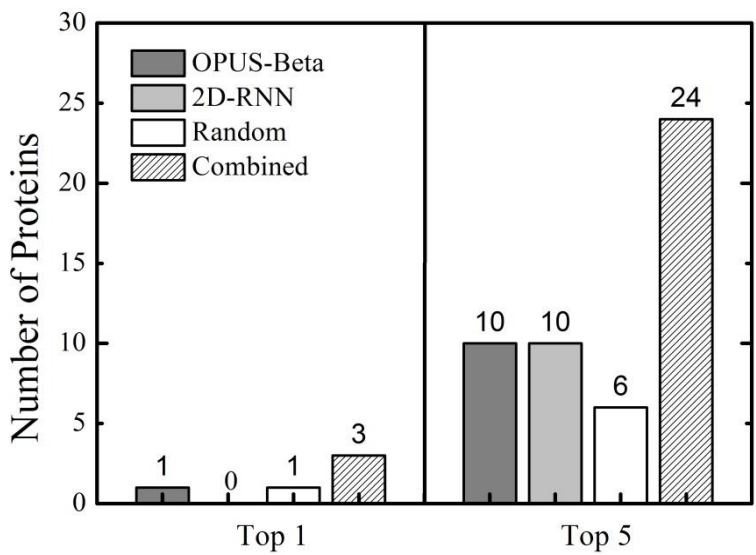

# Figure Legends

**Figure 1.** Schematic illustration of the five different self-packing patterns for the self-packing energy term: (a) one parallel partner; (b) one antiparallel partner; (c) two parallel partners; (d) two antiparallel partners; (e) one parallel partner and one antiparallel partner.

**Figure 2.** Schematic illustration of different types of pairwise packing patterns between two β-residues. (a) All possible pairwise residue-residue contacts in two antiparallel strands, including four types of inter-strand pairs and two types of intra-strand pairs. AA', a hydrogen-bond-involving inter-strand pair (illustrated by $L_{i-1,j+1}$ based on the diagram); AB', a hydrogen-bond-involving residue interacting with the next hydrogen-bond-involving residue on the opposite strand (illustrated by $L_{i-1,j-1}$ based on the diagram); aa', a non-hydrogen-bond-involving inter-strand pair (illustrated by $L_{i,j}$ based on the diagram); ab', a non-hydrogen-bond-involving residue interacting with the next non-hydrogen-bond-involving residue on the opposite strand (illustrated by $L_{i,j-2}$ based on the diagram); ab, a non-hydrogen-bond-involving intra-strand pair (illustrated by $L_{i,i+2}$ based on the diagram); AB, a hydrogen-bond-involving intra-strand pair (illustrated by $L_{i-1,i+1}$ based on the diagram). (b) All possible pairwise residue-residue contacts in two parallel strands, including three types of inter-strand pairs and two types of intra-strand pairs. Aa, a hydrogen-bond-involving residue interacting with a non-hydrogen-bond-involving residue (illustrated by $L_{i-1,j-1}$ based on the diagram); Ab, a hydrogen-bond-involving residue interacting with the next non-hydrogen-bond-involving residue on the opposite strand toward the C terminus (illustrated by $L_{i-1,j+1}$ based on the diagram); Ba, a hydrogen-bond-involving residue interacting with the next non-hydrogen-bond-involving residue on the opposite strand toward the N terminus (illustrated by $L_{i+1,j-1}$ based on the diagram); ab, a non-hydrogen-bond-involving intra-strand pair (illustrated by $L_{j-1,j+1}$ based on the diagram); AB, a hydrogen-bond-involving intra-strand pair (illustrated by $L_{i-1,i+1}$ based on the diagram).

**Figure 3.** Self-packing energy values for amino-acid residues involved in five different β-sheet packing patterns (also see Fig.1): A, one parallel partner; B, one antiparallel partner; C, two parallel partners; D, two antiparallel partners; E, one parallel partner and one antiparallel partner. Negative energy values indicate more favorable packing energetics, while positive values indicate less favorable packing energetics.

**Figure 4.** The performance of different kinds of potentials on decoy set I. X-axis: the native contact pattern was ranked top-1 (left) and within top-5 (right). Y-axis: the number of proteins that reached the indicated performance (X-axis) out of a total of 100 proteins tested. The 2D-NRR potential is derived from the contact map produced by 2D-RNN method, in which we used a threshold of 8 Å. As a control, RANDOM provides the results ranked by random selections.

**Figure 5.** The performance of different kinds of potentials on decoy set II. X-axis: the native contact pattern was ranked top-1 (left) and within top-5 (right). Y-axis: the number of proteins that reached the indicated performance (X-axis) out of a total of 100 proteins tested. The 2D-NRR potential is derived from the contact map produced by 2D-RNN method, in which we used a threshold of 8 Å. As a control, RANDOM provides the results ranked by random selections.

**Figure 6.** The performance of different kinds of potentials on combined decoy from set I and II. X-axis: the native contact pattern was ranked top-1 (left) and within top-5 (right). Y-axis: the number of proteins that reached the indicated performance (X-axis) out of a total of 100 proteins tested. The 2D-NRR potential is derived from the contact map produced by 2D-RNN method, in which we used a threshold of 8 Å. As a control, RANDOM provides the results ranked by random selections.